\documentclass[10pt, conference]{IEEEtran}
\usepackage[utf8x]{inputenc} 
\usepackage{graphicx}
\usepackage{float}
\usepackage{scalefnt} 
\usepackage[table,xcdraw]{xcolor}
\usepackage{url}
\usepackage{multirow}
\usepackage{scalefnt} 
\usepackage{verbatim}
\usepackage{booktabs}

\begin{document}

\newcommand\blfootnote[1]{%
  \begingroup
  \renewcommand\thefootnote{}\footnote{#1}%
  \addtocounter{footnote}{-1}%
  \endgroup
}

\title{Evaluating Convolutional Neural Networks for COVID-19 classification in chest X-ray images}

\newif\iffinal
\finaltrue
\newcommand{\jemsid}{34}

\iffinal
\author{
    \IEEEauthorblockN{Leonardo Gabriel Ferreira Rodrigues, Larissa Ferreira Rodrigues, Danilo Ferreira da Silva, João Fernando Mari}
		\IEEEauthorblockA{Instituto de Ciências Exatas e Tecnológicas\\
		Universidade Federal de Viçosa - UFV\\
		Caixa Postal 22 - 38.810-000 - Rio Paranaíba - MG - Brasil\\
		Email: \{leonardo.g.rodrigues, larissa.f.rodrigues, danilo.f.silva, joaof.mari\}@ufv.br}
}
\else
  \author{WVC paper ID: \jemsid \\ \\ \\ \\}
\fi
\maketitle

\begin{abstract}
\blfootnote{\color{red} The definite version of this paper was published by SBC-OpenLib as part of the WVC 2020 proceedings. DOI: \url{https://doi.org/10.5753/wvc.2020.13480}} 
    Coronavirus Disease 2019 (COVID-19) pandemic rapidly spread globally, impacting the lives of billions of people. The effective screening of infected patients is a critical step to struggle with COVID-19, and treating the patients avoiding this quickly disease spread. The need for automated and scalable methods has increased due to the unavailability of accurate automated toolkits. Recent researches using chest X-ray images suggest they include relevant information about the COVID-19 virus. Hence, applying machine learning techniques combined with radiological imaging promises to identify this disease accurately. It is straightforward to collect these images once it is spreadly shared and analyzed in the world. This paper presents a method for automatic COVID-19 detection using chest X-ray images through four convolutional neural networks, namely: AlexNet, VGG-11, SqueezeNet, and DenseNet-121. This method had been providing accurate diagnostics for positive or negative COVID-19 classification. We validate our experiments using a ten-fold cross-validation procedure over the training and test sets. Our findings include the shallow fine-tuning and data augmentation strategies that can assist in dealing with the low number of positive COVID-19 images publicly available. The accuracy for all CNNs is higher than 97.00\%, and the SqueezeNet model achieved the best result with 99.20\%.
\end{abstract}


\IEEEpeerreviewmaketitle

\section{Introduction}\label{intro}  
Coronavirus Disease 2019 (COVID-19) caused by a novel coronavirus, officially named Severe Acute Respiratory Syndrome Coronavirus 2 (SARS-CoV-2) \cite{WHO2020}, is a pandemic that first emerged in the Chinese city of Wuhan, and rapidly spread to other countries, affecting Italy, Iran, Spain, Brazil, Russia, India, and United States severely. All countries affected by COVID-19 imposed a nationwide lockdown in an attempt to slow the spread of the virus, causing a profound overall impact on the lives of billions of people from a health, safety, and economic perspective \cite{Zhou2020} \cite{Onder2020}. The COVID-19 can cause illness to the respiratory system leading to inflammation of the lungs and pneumonia \cite{Guan2020}. There is no known specific therapeutic drugs or vaccine for COVID-19, and the impact in the healthcare system is also high due to the number of people that needs intensive care unit (ICU) admission and breathing machine for long periods \cite{Grasselli2020}.

The most common test technique currently used for COVID-19 diagnosis is the reverse transcription-polymerase chain reaction (RT-PCR). However, considering the difficulties of distributing the kits and collecting the samples and the waiting time for results, auxiliary diagnostics methods are welcome to assist the medical team decision making. In this context, the development of computer-aided diagnosis systems based on machine learning is essential and widely applied in several fields of medicine \cite{NANNI2018} \cite{KERMANY2018} \cite{RODRIGUES2020}. 

Early studies demonstrated that many patients infected with COVID-19 present abnormalities in chest X-ray images \cite{Fang2020} \cite{Ai2020} \cite{Rubin2020}. These images can be easily collected, shared, and analyzed around the world. Moreover, the task of COVID-19 identification is not easy, and the specialist reviewing the chest X-ray needs to look for white patches in the lungs, i.e., air sacs filled with pus or water. However, these white patches can also be confused with diseases such as tuberculosis, bronchitis, and other types of pneumonia caused by different viruses or bacteria.

In this study, we aim to explore the identification of COVID-19 using chest X-ray images due to its reduced cost, fast result, and general availability. Our principal goal is to reach the best possible identification rate among COVID-19 and other types of pneumonia. We applied a pure deep learning approach comparing four Convolutional Neural Networks (CNNs): AlexNet, VGG-11, SqueezeNet and DenseNet-121, and we evaluated the performance using a k-fold cross-validation procedure over the training and test sets. Moreover, we carried a confusion matrix analysis to measure accuracy, precision, recall, and F1-score indices.

Furthermore, the chest X-ray image dataset used in this work contains a few positive images on the COVID-19. To deal with this, we applied Shallow Fine-Tuning (SFT) training and data augmentation based on random rotation and shifting to balance the class distribution and the performance of the classification. In addition, our approach is fast and simple producing high performing system. As far as we know, our result is the best obtained for COVID-19 identification in chest X-ray images. We believe that our proposed method can contribute to future researches intended to help healthcare workers to identify COVID-19 and to manage patient's conditions.



The remaining of this paper is organized as follows: Section \ref{sec:related-work}  surveys related work; Section \ref{sec:methods} describes the material and methods; In Section \ref{sec:results} we present and discuss the results obtained. Finally, conclusions and future work are presented in Section \ref{sec:conclusion}.

\section{Related Work}\label{sec:related-work}  
The COVID-19 has been attracting much attention from the image analysis research community due to its severity. In this sense, Narin et al. \cite{Narin2020} compared three different CNN architectures (ResNet50, Inception-V3, and InceptionResNetV2) to identify COVID-19 in chest X-ray images. They used a dataset composed of fifty COVID-19 images taken from the open-source GitHub repository shared by Dr. Joseph Cohen \cite{cohen2020covid} and fifty healthy lung images from Kaggle repository ``Chest X-Ray Images (Pneumonia)'' \cite{mooney2018chest}. The ResNet-50 obtained the best result achieving an accuracy of 98\%. 

Hemdan et al. \cite{Hemdan2020} proposed a COVIDX-Net composed of seven popular CNN models and used the same dataset considered by \cite{Narin2020} and achieved 90\% in terms of accuracy. However, only 25 samples of COVID-19 positive and 25 samples of negative images were considered.  

Sethy and Behera \cite{Sethy2020} also considered the same dataset of \cite{Hemdan2020}. Their study states that the ResNet-50 as a feature extractor and Support Vector Machine (SVM) classifier provided the best performance obtained an accuracy of 95.38\%.

Wang and Wong \cite{Wang2020} proposed a COVID-Net architecture, an open-source CNN created to detect COVID-19 on chest X-ray images. The authors used a dataset created exclusively to support COVID-Net experimentation, which obtained 93.3\% accuracy in classifying normal, non-COVID pneumonia, and COVID-19 classes.

Apostolopoulos and Mpesiana \cite{Apostolopoulos2020} adopted different pre-trained network architectures to address the task of classification of COVID-19 in chest X-ray images and achieved 96.78\% of accuracy with MobileNet v2 model. 

Khan et al. \cite{Khan2020} proposed the CoroNet deep CNN with 71 layers, inspired by Xception (Extreme Inception) and trained on the ImageNet dataset \cite{ImageNet}. According to the authors, the CoroNet was evaluated using a dataset not publicly available for download and achieved an average accuracy of 89.60\% for the COVID-19 identification.  

Ozturk et al. \cite{Ozturk2020} proposed an approach for early detection of COVID-19 cases using the DarkNet model as a classifier YOLO object detection system and obtained an accuracy of 98.08\% for binary classes and 87.02\% for multi-class cases. Ucar and Korkmaz \cite{Ucar2020} proposed a method based on SqueezeNet \cite{Forrest2016} architecture with Bayes optimization and achieved 98.30\% of accuracy.

Pereira et al. \cite{Pereira2020} utilized texture descriptors, fusion techniques and a pre-trained Inception-V3 model to identify COVID-19 obtained 89.00\% in terms of F1-score. However, the validation methodology used in \cite{Ucar2020} and \cite{Pereira2020} is a simple hold-out technique that has a certain probability of building biased sets, which may achieve abnormal accuracy results, mainly in small datasets. 

The automated classification of COVID-19 in X-ray images is a hot topic nowadays due to the growing pandemic, and new works are emerging every day. In contrast to the previous works, we explore training based on SFT, data augmentation strategy, and our approach is a promising alternative by delivering a simple and efficient that allows achieved better results with a low computational cost.

\section{Material and Methods} \label{sec:methods}

\begin{figure}[!htbp]
    \begin{center}
    	\includegraphics[width=0.94\columnwidth]{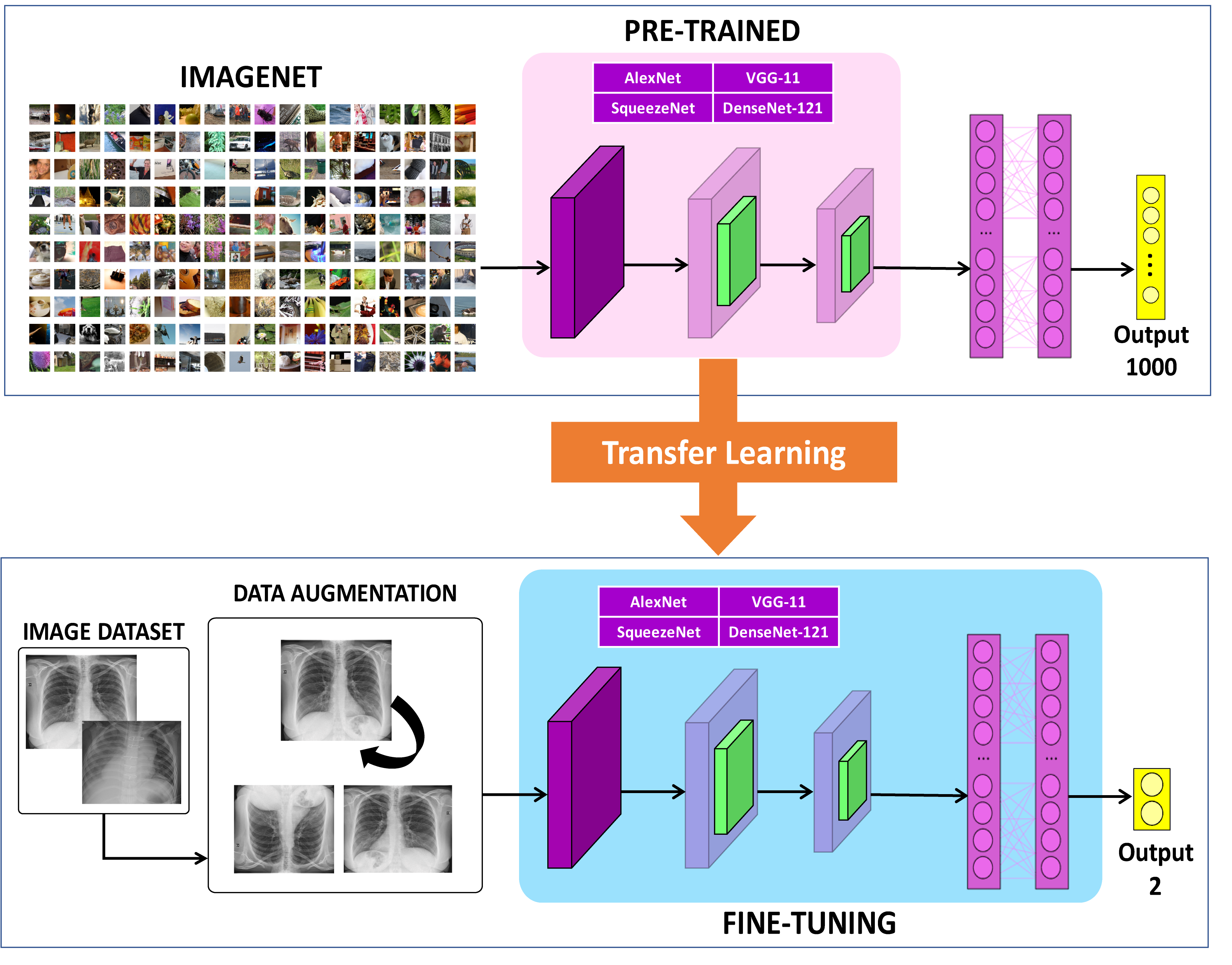}
    \end{center}
    \caption{Steps of proposed method.}
	\label{fig:steps}
\end{figure}

The main goal of this paper is to evaluate the performance of different architectures of CNNs to classify COVID-19 in chest X-ray images. More precisely, we find the best possible identification rate among COVID-19 and other types of pneumonia. Fig. \ref{fig:steps} illustrates the steps of the methodology adopted here.

\subsection{Image dataset}

The images used in this work were obtained from two datasets of chest X-ray images. The first dataset contains 108 images of COVID-19 positive and was taken from the GitHub repository shared by Dr. Joseph Cohen, at the University of Montreal \cite{cohen2020covid} (last accessed April 10, 2020). We selected 299 images of COVID-19 negative, corresponding to 20\% of viral pneumonia images selected randomly from the Chest X-Ray Images (Pneumonia) dataset available in Kaggle repository \cite{mooney2018chest}. Note that only about 20\% of the viral pneumonia was selected in order to avoid imbalance between the classes or bias the classification performance. 

\begin{figure}[!htb]
	\includegraphics[width=0.88\columnwidth]{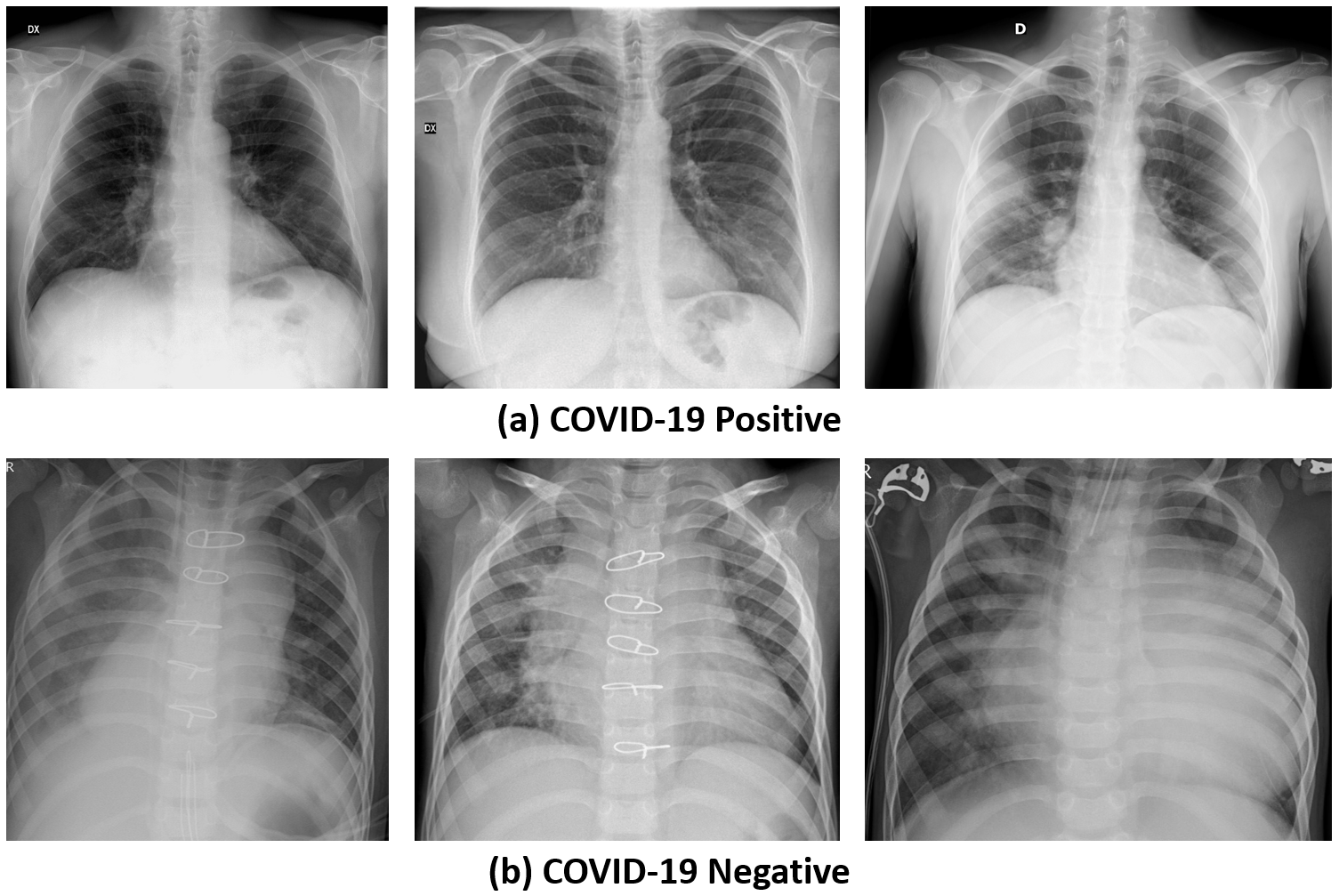}
	\centering
	\caption{Examples of image instance for each class.}
	\label{fig:dataset}
\end{figure}

The information about images is summarized in Table \ref{tab:dataset}. To illustrate the dataset resulting from the combination of the two datasets previously mentioned, sample images from each class are presented in Fig. \ref{fig:dataset}. 

\begin{table}[!htb]
\renewcommand{\arraystretch}{1.4}
\centering
\caption{Distribution of the chest X-ray images.}
\label{tab:dataset}
\resizebox{\columnwidth}{!}{
\begin{tabular}{lcc} \hline
\multicolumn{1}{c}{\textbf{COVID-19}} & \textbf{Samples} & \textbf{Source}                                                              \\ \hline
Positive                              & 108              & \begin{tabular}[c]{@{}c@{}}GitHub\\ (Dr. Joseph Cohen) \cite{cohen2020covid}\end{tabular}          \\ \hline
Negative                              & 299              & \begin{tabular}[c]{@{}c@{}}Kaggle\\ (X-ray images of Pneumonia) \cite{mooney2018chest}\end{tabular} \\ \hline
\textbf{Total}                        & \textbf{407}     & \multicolumn{1}{l}{}              \\ \hline                                          
\end{tabular}
}
\end{table}

\subsection{Pre-processing}

All images were resized to 224 $\times$ 224 pixels based on bilinear interpolation. The resize allows adapting each image for the input of the CNN architectures used in this work.

As one of the main obstacles in this study is the lack of images, we applied data augmentation strategy to increase the training data artificially without introducing labeling costs \cite{Krizhevsky2012}. All training images were augmented by using vertical and horizontal flips, and rotating each of the original images around its center through randomly chosen angles of between -10$^{\circ}$ and 10$^{\circ}$.  

\subsection{Convolutional Neural Networks (CNNs)} \label{subsec:CNN}

Convolutional Neural Networks (CNNs) are a multi-stage image classification technique that incorporates spatial context and weight sharing between pixels in order to extract high-level hierarchical representations of the data \cite{Goodfellow2016} \cite{Ponti2017}. Thus, CNN is able to extract features during training. In this work, four CNNs architectures are tested:  AlexNet \cite{Krizhevsky2012}, VGG-11 \cite{Simonyan2014}, SqueezeNet \cite{Forrest2016}, and DenseNet-121 \cite{Huang2017}. All CNNs were selected based on their success in previous image classification tasks.

The AlexNet was the champion of ImageNet Large Scale Visual Recognition Challenge (ILSVRC) 2012 \cite{ImageNet}. This CNN consists of five convolutional layers, three max-pooling layers, two fully connected layers with a final softmax layer. In order to reduce overfitting, AlexNet uses dropout connections and REctified Linear Unit (ReLU) activation function \cite{Krizhevsky2012}.
    
The VGG won the identification and classification tasks in the ILSVRC 2014. In order to reduce the computational cost, this CNN adopted sequences of convolutional filters of size 3$\times$3 \cite{Simonyan2014}. In this paper, we use the VGG-11 architecture with batch normalization, due to its simplicity and robustness. It is important to note that batch normalization is very effective to overcome the challenges of deep training \cite{Ioffe2015}.
    
SqueezeNet is a compact CNN with approximately 50 times fewer parameters than AlexNet model. It is composed of a stand-alone convolution layer followed by eight fire modules and a final convolution layer \cite{Forrest2016}. Each fire module contains only a filter of size 1$\times$1 inputting into an expanded layer composed by convolutional filters of size 1$\times$1 and 3$\times$3. In this way, the modules are able to perform the same functions of fully connected and dense layers. 
    
DenseNet-121 architecture uses dense blocks to concatenate a number of convolutional layers reducing the number o features through average pooling. In the present study, we use the DenseNet with 121 layers: one initial convolutional layer followed by max-pooling, 116 convolutional layers followed by batch normalization, and ReLU functions, interpolated with three transition blocks, and a last average pooling before the start of fully connected layers \cite{Huang2017}.

\subsection{Shallow Fine-Tuning}\label{subsec:SFT}

The Shallow Fine-Tuning (SFT) \cite{Tajbakhsh2016} \cite{Izadyyazdanabadi2018} is a training strategy based on the concept of transfer learning and is suitable for small data sets. This approach is used to train deep learning models in which the network is pre-trained for a classification task using a huge dataset such as ImageNet \cite{ImageNet}. 

The weights in all convolutional layers are initialized with the corresponding values from the pre-trained model. These layers are considered more general and retain information about texture, color, and shape. SFT performs fine-tuning only in the last fully connected layer, which is more specialized. Usually, this strategy is the most common allowing weights of the last layers to adapt to the classification problem.


\subsection{Training strategy}
The training of the CNNs models is defined as an optimization problem in order to optimize the quality of the prediction. In this work, we considered the objective function as the cross-entropy defined by $\mathcal{L}(W)$. Equation \ref{eq:loss_function} show that $\mathcal{L}(W)$ is computed over a set of training samples $X_{j}$ considering the tuned weights $W$, parameters $f(x_{j})$, and the known classes $y_{j}$, where $j$ represents the classes COVID-19 positive and negative.

\begin{equation}
    \label{eq:loss_function}
    \centering
    \mathcal{L}(W) = \frac{1}{n}\sum_{j=1}^{N}\ell (y_{j}, f(x_{j}; W))
\end{equation}

In this way, to minimize $\mathcal{L}(W)$, we applied the Stochastic Gradient Descent (SGD) \cite{Lecun1998} optimization algorithm with momentum of 0.9, learning rate of 0.001, and batch size of eight. All CNNs were trained for 30 epochs.

\subsection{Evaluation methodology}

Due to the very small number of positive images of COVID-19 available,  we have decided not to perform hyperparameter optimization nor early stop strategy. These procedures require a validation set, which further reduces the number of images available for testing the models. For this reason, all CNN models were trained and tested using the stratified $k$-fold cross-validation method \cite{Devijver1982}. All chest X-ray images were randomly partitioned into ten folds. Then, the model is trained with $k-1$ folds and tested on the remaining fold. The training and testing procedures are repeated k times, alternating the testing folds. Thus, we guarantee that each image will participate in the training process ($k - 1$ times) and will also be part of the test group (1 time). Finally, the results from the $k$ testing sets are averaged to produce a single and trustworthy estimation.

The metrics used to assess the classification performance include accuracy, precision, recall, and F1-score indices. All indices are based on the number of true positives (TP), true negatives (TN), false positives (FP), and false-negative (FN) classifications obtained from the confusion matrix \cite{Duda2000}. Also, we measure the standard deviation in order to assess the confidence of results, where smaller values represent high-reliability.

\begin{itemize}
    \item Accuracy: is the ratio between the correct classifications and total samples (Eq.\ref{eq:acc}).
    
    \begin{equation}
    \label{eq:acc}
        Accuracy = \frac{TP + TN}{TP + TN + FP + FN} 
    \end{equation}
    
    \item Precision: is the ratio between TP and the total of positives classification (Eq. \ref{eq:prc})
    
    \begin{equation}
    \label{eq:prc}
        Precision = \frac{TP}{TP + FP}  
    \end{equation}

    \item Recall: is the harmonic average of recall and precision (Eq. \ref{eq:rec}).

    \begin{equation}
    \label{eq:rec}
        Recall = \frac{TP}{TP + FN}  
    \end{equation}

    \item F1-Score: is the weighted average of the precision and recall (Eq. \ref{eq:fsc})
 
    \begin{equation}
    \label{eq:fsc}
        F1\!\!-\!Score  = 2 \times \frac{Precision \times Recall}{Precision + Recall}
    \end{equation}   
    
\end{itemize}\smallskip

Also, we used the Receiver Operating Characteristic (ROC), and the Area Under ROC (AUC) as a reliable classification performance measure of all possible classification thresholds.


\section{Results and Discussion}\label{sec:results}

All experiments were programmed using Python (version 3.6) and PyTorch (version 1.4) deep learning framework \cite{PyTorch}. This study investigated the performance of four CNNs architectures to classify chest X-ray images on COVID-19 positive and COVID-19 negative (pneumonia) classes. 

\subsection{Comparison of architectures}
One of the most challenges in training CNNs architectures for classification tasks is the lack of large enough datasets for adjust a large number of model parameters. We adopted SFT fine-tuning training strategy to overcome this problem, instead of training the models from scratch (as described in Section \ref{subsec:SFT}). Thus, the weights in the initial layers of CNN (simpler features) were kept, but the weights of the deeper layers (more specialized features) were adapted to the problem of classifying X-ray images generating greater specialization in the deep layers. It is important to mention that number of positive COVID-19 images in the public repository is very small. To overcome this issue, we applied data augmentation strategies aim to increase the size of the training set.

\begin{figure}[!htbp]
	\includegraphics[width=1\columnwidth]{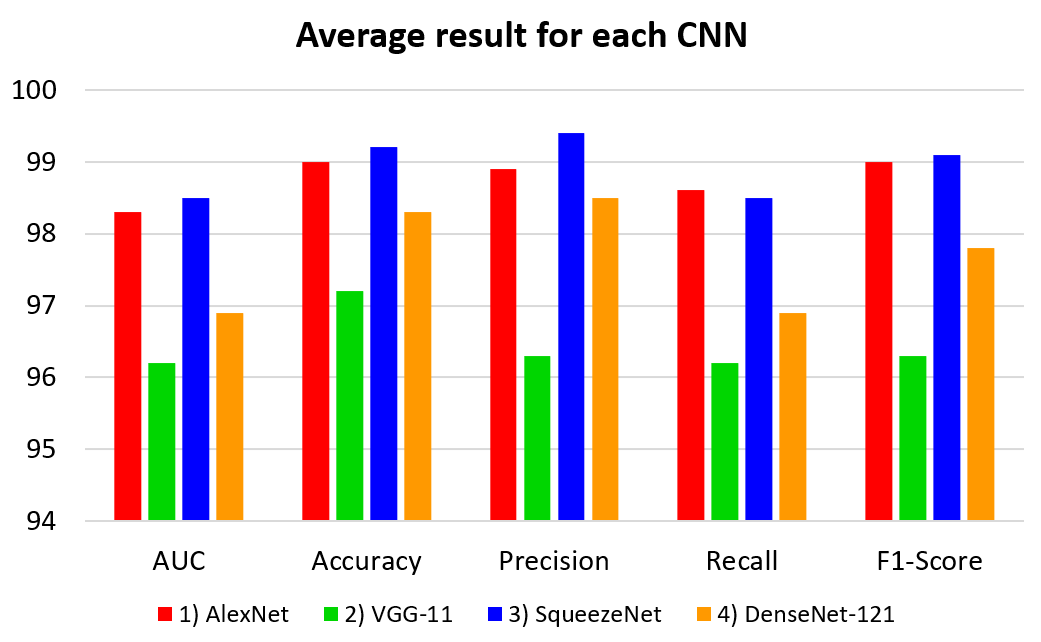}
	\centering
	\caption{10-fold average values of the performance measures for each CNN model. 1) AlexNet; 2) VGG-11; 3) SqueezeNet; and 4) DenseNet-121.}
	\label{fig:avg}
\end{figure}

Table \ref{tab:results} presents the average classification performance considering the results of accuracy, precision, recall and F1-score indices concerning each CNN evaluated. The chart in Fig. \ref{fig:avg} illustrate the variation of each performance values based on the results presented in Table \ref{tab:results}. Interestingly, SqueezeNet achieved the best performance, followed by AlexNet, DenseNet-121, and VGG-11. Also, SqueezeNet is recognized as having a low computational cost and designed for embedded systems. Therefore, the result suggested that our proposed could be used to evaluate chest X-ray images using mobile devices. Moreover, the standard deviation value obtained for each CNN was small and indicates that our results are reliable.

\begin{table*}[!htbp]
\renewcommand{\arraystretch}{1.3}\scalefont{1.2}
\scalefont{1.2}
\centering
\caption{10-fold average values and standard deviation of the performance measures for each CNN model.}
\label{tab:results}
\begin{tabular}{lccccc}
    \midrule
    \multicolumn{1}{c}{\textbf{CNN}} & \textbf{AUC (\%)} & \textbf{Accuracy (\%)} & \textbf{Precision (\%)} & \textbf{Recall (\%)} & \textbf{F1-Score (\%)} \\
    \midrule
    AlexNet & 98.30 $\pm$ 0.02 & 99.00 $\pm$ 0.01 & 98.90 $\pm$ 0.02 & 98.60 $\pm$ 0.02 & 99.00 $\pm$ 0.01 \\
    VGG-11 & 96.20 $\pm$ 0.04 & 97.20 $\pm$ 0.03 & 96.30 $\pm$ 0.04 & 98.60 $\pm$ 0.04 & 99.00 $\pm$ 0.04 \\
    SqueezeNet & 98.50 $\pm$ 0.02 & 99.20 $\pm$ 0.01 & 99.40 $\pm$ 0.01 & 98.50 $\pm$ 0.02 & 99.10 $\pm$ 0.01 \\
    DenseNet-121 & 96.90 $\pm$ 0.03 & 98.30 $\pm$ 0.02 & 98.50 $\pm$ 0.01 & 96.90 $\pm$ 0.03 & 97.80 $\pm$ 0.02 \\
    \midrule
    \end{tabular}%
\end{table*}

\begin{table*}[!htbp]
\renewcommand{\arraystretch}{1.4}\scalefont{1.3}
\scalefont{1.2}
\centering
\caption{10-fold values of confusion matrix for each CNN model.}
\label{tab:matrix}
\resizebox{\columnwidth}{!}{
\begin{tabular}{lccllcc}
                                        & \multicolumn{2}{c}{\textbf{AlexNet}}                                            &                       &                                        & \multicolumn{2}{c}{\textbf{VGG-11}}                                             \\ \cline{2-3} \cline{6-7} 
\multicolumn{1}{l|}{}                   & \multicolumn{1}{c|}{\textbf{Positive}} & \multicolumn{1}{c|}{\textbf{Negative}} &                       & \multicolumn{1}{l|}{}                  & \multicolumn{1}{c|}{\textbf{Positive}} & \multicolumn{1}{c|}{\textbf{Negative}} \\ \cline{1-3} \cline{5-7} 
\multicolumn{1}{|l|}{\textbf{Positive}} & \multicolumn{1}{c|}{105}               & \multicolumn{1}{c|}{3}                 & \multicolumn{1}{l|}{} & \multicolumn{1}{l|}{\textbf{Positive}} & \multicolumn{1}{c|}{102}               & \multicolumn{1}{c|}{6}                 \\ \cline{1-3} \cline{5-7} 
\multicolumn{1}{|l|}{\textbf{Negative}} & \multicolumn{1}{c|}{1}                 & \multicolumn{1}{c|}{298}               & \multicolumn{1}{l|}{} & \multicolumn{1}{l|}{\textbf{Negative}} & \multicolumn{1}{c|}{6}                 & \multicolumn{1}{c|}{293}               \\ \cline{1-3} \cline{5-7} 
                                        & \multicolumn{1}{l}{}                   & \multicolumn{1}{l}{}                   &                       &                                        & \multicolumn{1}{l}{}                   & \multicolumn{1}{l}{}                   \\
                                        & \multicolumn{2}{c}{\textbf{SqueezeNet}}                                         &                       &                                        & \multicolumn{2}{c}{\textbf{DenseNet-121}}                                       \\ \cline{2-3} \cline{6-7} 
\multicolumn{1}{l|}{}                   & \multicolumn{1}{c|}{\textbf{Positive}} & \multicolumn{1}{c|}{\textbf{Negative}} &                       & \multicolumn{1}{l|}{}                  & \multicolumn{1}{c|}{\textbf{Positive}} & \multicolumn{1}{c|}{\textbf{Negative}} \\ \cline{1-3} \cline{5-7} 
\multicolumn{1}{|l|}{\textbf{Positive}} & \multicolumn{1}{c|}{105}               & \multicolumn{1}{c|}{3}                 & \multicolumn{1}{l|}{} & \multicolumn{1}{l|}{\textbf{Positive}} & \multicolumn{1}{c|}{102}               & \multicolumn{1}{c|}{6}                 \\ \cline{1-3} \cline{5-7} 
\multicolumn{1}{|l|}{\textbf{Negative}} & \multicolumn{1}{c|}{0}                 & \multicolumn{1}{c|}{299}               & \multicolumn{1}{l|}{} & \multicolumn{1}{l|}{\textbf{Negative}} & \multicolumn{1}{c|}{1}                 & \multicolumn{1}{c|}{298}               \\ \cline{1-3} \cline{5-7} 
\end{tabular}
}
\end{table*}

As in each fold the images in the testing sets do not repeat, we consolidate the confusion matrices from each fold by adding the values from each confusion matrix. Therefore, the confusion matrix presented in Table \ref{tab:matrix} summarizes results for all ten folds, and presents the prediction for all images in the dataset.

The confusion matrices of each CNN model allow observing several aspects of the classification problem investigated in this work. Note that for all CNNs, the COVID-19 positive and COVID-19 negative is well identified. In particular, AlexNet and SqueezeNet models were able to classify the most positive cases of COVID-19 correctly. The results suggest that models have been able to preserve in the feature maps important information about visual patterns of diagnostic positive. It is important to mention that number of positive COVID-19 images in the public repository is very small. In this study, to compare the performance of different CNN architectures, we focused to obtain reliable and trustfully results. 

With the lack of data, we still cannot recommend these methods as a diagnostic aid system, but our results support that the use of CNN models is a promising technique to assist the early diagnosis of COVID-19 in conjugation with other standard tests. However, the number of available images tends to grow as studies advance, and with more accurate researches, it will be possible to understand the capability of CNNs in helping detecting COVID-19.

\subsection{Comparison with literature}

The best result achieved in this study in terms of accuracy and F1-score is compared with other state-of-art work in the literature. The best result in our work was obtained with SqueezeNet, trained with SFT using augmented data, which scored 99.20\% of accuracy and 99.10\% of F1-score (as shown in Table \ref{tab:results}). The best results reported in the literature are presented in Table \ref{tab:comparison} for the same COVID-19 dataset. It can be seen that our best score is upper to the best state-of-the-art technique reported in the literature.

\begin{table}[!htb]
\renewcommand{\arraystretch}{1.3}
\centering
\caption{Highest accuracy of other classification methods using the COVID-19 dataset from GitHub \cite{cohen2020covid}.}
\label{tab:comparison}
\resizebox{\columnwidth}{!}{
\begin{tabular}{lc} \midrule
\multicolumn{1}{c}{\textbf{Method}}              & \textbf{Accuracy (\%)}  \\ \midrule
Narin et al. \cite{Narin2020}                                     & 98.00                   \\
Hemdan et al. \cite{Hemdan2020}                                   & 90.00                   \\
Sethy and Behera \cite{Sethy2020}                                & 95.38                   \\
Wang and Wong \cite{Wang2020}                                   & 93.30                   \\
Apostopoulos and Mpsiana \cite{Apostolopoulos2020}                         & 96.78                   \\
Khan et al. \cite{Khan2020}                                      & 89.60                      \\
Ozturk et al. \cite{Ozturk2020}                                    & 98.08                   \\
\textbf{Our work (SqueezeNet + SFT + data aug.)} & \textbf{99.20}          \\
                                                 & \multicolumn{1}{l}{}    \\ \midrule
                                                 & \textbf{F1-Score (\%)} \\ \midrule
Pereira et al. \cite{Pereira2020}                                   & 89.00                   \\
\textbf{Our work (SqueezeNet + SFT + data aug.)} & \textbf{99.10}     \\ \midrule  
\end{tabular}
}
\end{table}

\section{Conclusion}\label{sec:conclusion}

The results presented in this paper point to a promising using of CNN models to classify COVID-19 cases based on chest X-ray images. We compared the performance of four CNN architectures to classify X-ray images in COVID-19 positive and negative (pneumonia) classes, and our training strategy consists of applying transfer learning with SFT and different data augmentation approaches. Our best result of 99.20\% was upper to the highest accuracy score presented in the literature; this result was obtained with SqueezeNet model. 

Although there are few COVID-19 positive chest X-ray images, we designed our experiments to minimize the effects of CNN training with small data sets. The training using fine-tuning and data augmentation aim to increase the classification rate, while the stratified k-fold cross validation allows more reliable results than simple hold-out. Now, it is necessary to wait for more images of positive COVID-19 to be available in order to train more reliable models that may confirm the positive perspectives demonstrated by this study.

The presented results open new opportunities towards better machine learning based on deep CNNs for automated detection of COVID-19 and developing of new computer-aided diagnosis applications. Moreover,  exciting opportunities and future works raise such as testing other CNN models, evaluating more data augmentation strategies, and applying some hyperparameter optimization and combining classifications techniques.


\section*{Acknowledgments}
We gratefully acknowledge the support of NVIDIA Corporation with the donation of the TITAN Xp GPU used for this research. We would like to thanks CAPES and FAPEMIG (Grant number CEX - APQ-02964-17) for the financial support. This study was financed in part by the Coordenação de Aperfeiçoamento de Pessoal de Nível Superior - Brasil (CAPES) - Finance Code 001.

\bibliography{referencias}

\end{document}